\documentclass[preprint,prl,superscriptaddress,prbib,11pt]{revtex4}
\usepackage{graphicx}
\usepackage{bm}
\usepackage{hyperref}
\usepackage{todonotes}
\usepackage{nicefrac}
\usepackage{bbm}
\usepackage[normalem]{ulem}
\usepackage{booktabs}
\usepackage{subfigure}
\usepackage{rotating}
\usepackage{color}
\usepackage{float}
\usepackage[switch,columnwise]{lineno}
\begin{document}

\title{Cascade of strongly correlated quantum states in a partially filled kagome flat band}

\author{Caiyun Chen}
\affiliation{Department of Physics, The Hong Kong University of Science and Technology, Clear Water Bay, Kowloon, Hong Kong SAR}
\affiliation{Institute for Advanced Study, The Hong Kong University of Science and Technology, Clear Water Bay, Kowloon, Hong Kong SAR}
\affiliation{These authors contributed equally}
\author{Jiangchang Zheng}
\affiliation{Department of Physics, The Hong Kong University of Science and Technology, Clear Water Bay, Kowloon, Hong Kong SAR}
\affiliation{These authors contributed equally}
\author{Yuman He}
\affiliation{Department of Physics, The Hong Kong University of Science and Technology, Clear Water Bay, Kowloon, Hong Kong SAR}
\affiliation{These authors contributed equally}
\author{Xuzhe Ying}
\affiliation{Department of Physics, The Hong Kong University of Science and Technology, Clear Water Bay, Kowloon, Hong Kong SAR}
\author{Soumya Sankar}
\affiliation{Department of Physics, The Hong Kong University of Science and Technology, Clear Water Bay, Kowloon, Hong Kong SAR}
\author{Luanjing Li}
\affiliation{Department of Physics, The Hong Kong University of Science and Technology, Clear Water Bay, Kowloon, Hong Kong SAR}
\author{Yizhou Wei}
\affiliation{Department of Physics, The Hong Kong University of Science and Technology, Clear Water Bay, Kowloon, Hong Kong SAR}
\author{Xi Dai}
\affiliation{Department of Physics, The Hong Kong University of Science and Technology, Clear Water Bay, Kowloon, Hong Kong SAR}
\author{Hoi Chun Po}
\email[]{hcpo@ust.hk}
\affiliation{Department of Physics, The Hong Kong University of Science and Technology, Clear Water Bay, Kowloon, Hong Kong SAR}
\author{Berthold J\"{a}ck}
\email[]{bjaeck@ust.hk}
\affiliation{Department of Physics, The Hong Kong University of Science and Technology, Clear Water Bay, Kowloon, Hong Kong SAR}

\date{\today}

\begin{abstract}
Coulomb interactions among charge carriers that occupy an electronic flat band have a profound impact on the macroscopic properties of materials. At sufficient strength, these interactions can give rise to captivating phenomena such as quantum criticality, Mott-Hubbard states, and unconventional superconductivity. The appearance of these characteristics sensitively depends on the number of electrons occupying the flat band states. Consequently, the search for new flat band materials with tunable charge carrier filling is a central research theme. In this work, we present experimental evidence obtained from scanning tunneling microscopy measurements for a cascade of strongly correlated states appearing in the partially occupied kagome flat bands of Co$_{1-x}$Fe$_x$Sn whose filling can be controlled by the Fe-doping level $x$. At elevated temperatures ($T\geq16\,K$), we detect a nematic electronic state across a broad doping range $0.05<x<0.25$. The comparison with model calculations reveals that strong Coulomb interactions ($U>100\,$meV) blend the states of two $3d$-orbital derived flat bands and impart a nematic order parameter. This state serves as the parent phase of a strongly correlated phase diagram: At lower temperatures $T<16\,$K, we find spectroscopic evidence for an orbital-selective Mott state enabled by the $3d$-orbital degeneracy of the Co atom. This state can only be detected in samples with ideal Fe doping ($x=0.17$) and descends into pseudogap phases upon electron and hole doping. At $T<8\,$K, the pseudogap phase evolves into another nematic low temperature state. Our observations demonstrate that the electronic ground state of a kagome flat band depends on the complex interplay between strong Coulomb repulsion, $3d$-orbital degeneracy, and flat band filling fraction at different temperatures. More broadly, our research establishes kagome materials as a unique platform to search for strongly correlated quantum states that arise in non-trivial flat bands and can be controlled by the filling fraction.
\end{abstract}

\maketitle

\section{Introduction}

Strong interactions among electric charge carriers can engender fascinating phenomena such as high-temperature superconductivity, non-Fermi liquids, and correlated insulating states that significantly influence the macroscopic material properties. The appearance of these characteristics is often connected with the suppression of the kinetic energy ($W$) of charge carriers occupying an electronic flat band (FB). As they move more slowly, the effect of the mutual Coulomb interaction $U$ can prevail and novel electronic ground states emerge at $W<U$. Prime examples of such class of materials are transition metal oxides~\cite{lee2006doping, keimer2015quantum}, whose strongly correlated phase diagram arises from strong Coulomb interactions within the spatially localized $d$-orbital of the transition metal ion, as well as the magnetic field-induced FB of quantum Hall systems~\cite{girvin2002quantum, stern2008anyons}. 

Recent attention has also been drawn to FBs arising within two-dimensional (2D) lattice structures, exemplified by the discovery of superconductivity, correlated insulators, and non-Fermi liquid behavior in moiré materials~\cite{cao2018unconventional, cao2018correlated, cao2020strange, jaoui2022quantum}. Extensive research efforts focus on understanding the phase diagram that results from the complex interplay between charge, orbital and spin degrees of freedom at different charge carrier densities. This physical richness motivates the search for other materials and concepts toward realizing strongly correlated states in electronic FBs, with the ultimate goal of finding a comparable repertoire of interesting quantum states extending to elevated temperatures.

Destructive quantum interference resulting from geometric frustration can also suppress the kinetic energy of electrons. This phenomenon can appear in materials based on the two-dimensional (2D) kagome lattice, which consists of corner-sharing triangles (see Fig.~\ref{fig:fig1}a). The destructive interference between electron wave functions localized on adjacent lattice sites gives rise to a flat band with electronic states that are strongly localized in real space at atomic length scales. When the kagome lattice is composed of transition metal atoms~\cite{ye2018massive,liu2018giant,yin2018giant,kang2020topological,meier2020flat,sales2021tuning, sales2022flat,yin2022topological,huang2022flat,sankar2023observation,chen2023visualizing}, the presence of substantial Coulomb interactions ($U\approx5-6\,$eV) within the $d$-orbital derived flat band states~\cite{kang2020topological} could promote the emergence of strongly correlated electronics states when the band is partially filled. Recent reports of non-Fermi liquid behavior and anomalous magnetism indicate a measurable effect of electronic interactions in kagome materials~\cite{chen2022emergent,huang2022flat,ye2024hopping,huang2024non,ekahana2024anomalous}. However, a fundamental challenge remains in achieving and controlling partial fillings of the kagome FBs that would allow for the observation of a filling-dependent phase diagram of distinct strongly correlated states.

CoSn has emerged as a highly promising candidate in this regard. It features two kagome FBs positioned just below the Fermi energy $E_{\rm F}$ as well as a set of $3d$-orbital derived dispersive bands~\cite{meier2020flat,liu2020orbital,kang2020topological,huang2022flat,jiang2023kagome}. Cobalt (Co) atoms arrange in a 2D kagome lattice within Co$_3$Sn planes, separated by alternating layers of stanene (Sn$_2$) composed of tin (Sn) (as seen in Fig.~\ref{fig:fig1}a). This stacking arrangement promotes the effective 2D nature of the electronic FBs~\cite{liu2020orbital,chen2023visualizing}. 
Previous studies employing angle-resolved photo-electron spectroscopy and scanning tunneling microscopy (STM) detected density of state maxima of the two FBs at energies of $E=-70\,$meV (`the upper flat band') and $E\approx -200\,$meV to $-300\,$meV (`the lower flat band'), respectively~\cite{kang2020topological,liu2020orbital}. These measurements also revealed indications of a small in-plane dispersion $W<100\,$meV and a renormalized quasiparticle velocity $v\leq10^4\,$m/s of the flat band electronic states. The on-site Coulomb repulsion among charge carriers within the $3d$-orbital derived flat band states ($U\approx5-6\,$eV in the unscreened limit)~\cite{kang2020topological} is thus expected to surpass the kinetic energy of the flat band electrons ($W/U\ll1$).

In this work, we report the observation of a cascade of strongly correlated states in the kagome FBs of CoSn. Chemical Fe-doping of epitaxially grown CoSn thin films lowers the Fermi energy and enables us to realize partial flat band occupations. Examining the low-energy electronic states across a wide Fe doping ($0\leq x\leq 0.24$) and temperature ($4\,\text{K}<T\leq25\,\text{K}$) range with the STM, we discover a rich phase diagram of strongly correlated states (shown in Fig.~\ref{fig:fig1}b), whose ground state sensitively depends on the Fe doping level.

\section{Realizing partially filled kagome flat bands in C\lowercase{o}S\lowercase{n}}

\begin{figure}[H]
    \centering
    \includegraphics[width=1\linewidth]{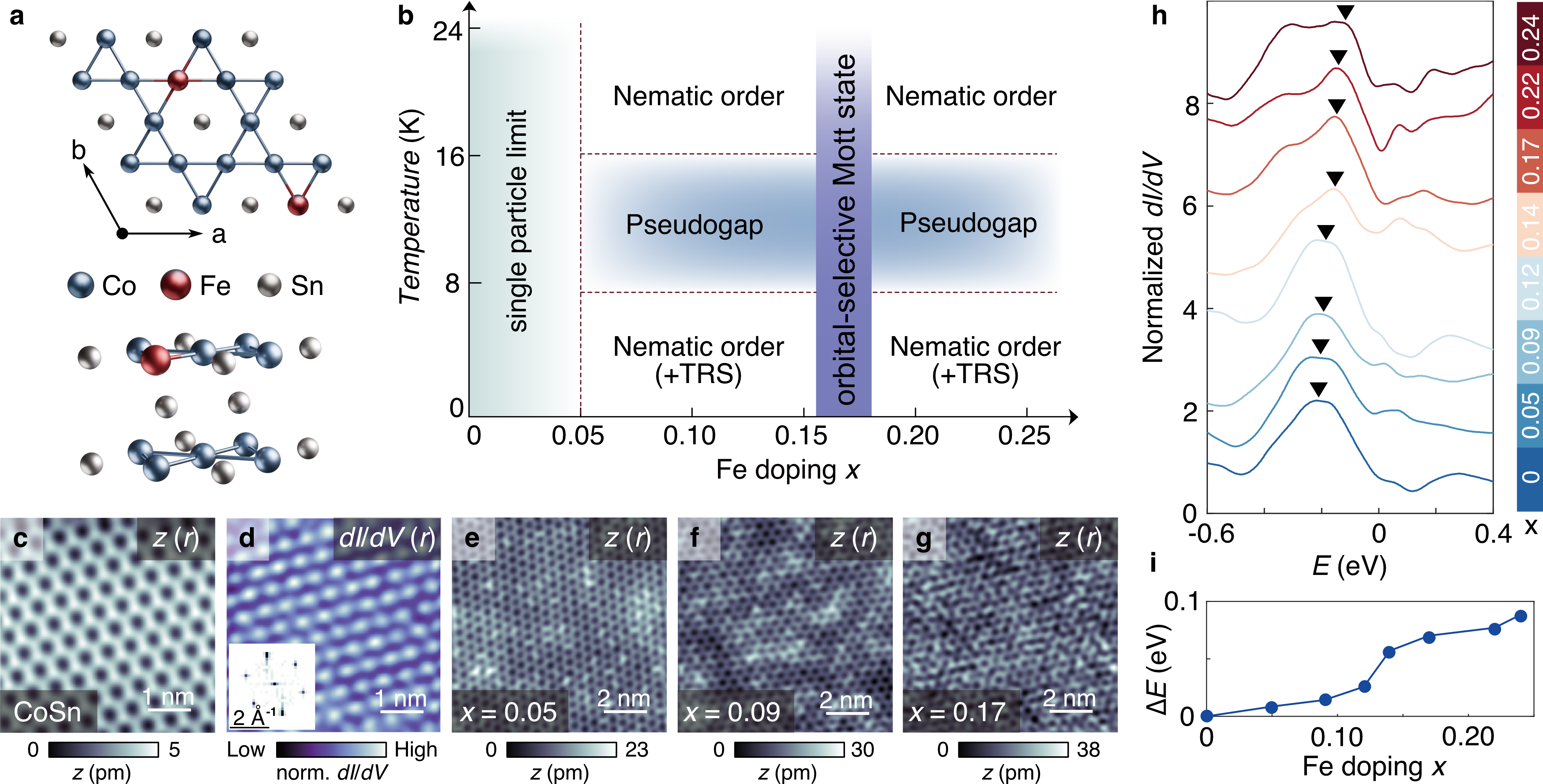}
    \caption{{\bf Cascade of strongly correlated states in the partially filled kagome flat bands in Fe-doped CoSn.} {\bf a,} Schematics of the in-plane lattice structure of Co$_{1-x}$Fe$_x$Sn (top) and the layer stacking in the out-of-plane direction (bottom). Doping with iron (Fe) substitutes cobalt (Co) atoms on the kagome lattice sites. {\bf b,} Temperature ($T$) and Fe doping ($x$) dependent phase diagram of interacting many body states in Co$_{1-x}$Fe$_x$Sn. Time-reversal-symmetry breaking (TRS) is inferred from Ref.~\cite{sales2021tuning} {\bf c,} STM topography on the surface of CoSn ($x=0$, bias voltage $V=100\,$mV, tunnel current set point $I=2.0\,$nA). {\bf d,} $dI/dV$ map recorded at $V=0\,$V ($I=4.0\,$nA, lock-in modulation voltage $V_{\rm m}=1.0\,$mV) in the same field of view as panel c. The inset shows the corresponding 2D fast Fourier transform (2D-FFT) where black and white color correspond to high and low amplitude, respectively. {\bf e}-{\bf g,} STM topographies recorded on samples with $x=0.05$, $0.09$, and $0.17$ ($V=1\,$V, $I=2.0\,$nA). {\bf h,} $dI/dV$ spectra recorded on the surface of samples with different indicated doping level $x$ at $T=4\,$K. The spectra are vertically offset for clarity ($V=0.8\,$V, $I=3.0\,$nA, $V_{\rm m}=10\,$mV). {\bf i,} Energetic shift $\Delta E$ of the flat band $dI/dV$ peak maximum, indicated by black triangle markers in panel h, plotted as a function of $x$.}
    \label{fig:fig1}
\end{figure}

We used molecular beam epitaxy (MBE) to realize a doping series of Co$_{1-x}$Fe$_x$Sn films (nominal thickness of 50\,nm) with different Fe doping levels $0\leq x\leq0.24$. Details of the MBE process are described in Ref.~\cite{chen2023visualizing} and the Methods section; the materials characterization is described in Sec.\,1 of the Supplementary Materials. The Vollmer-Weber growth mode results in flat-top islands of few hundreds nanometers in diameter, whose surface is commonly terminated by a Co$_3$Sn layer with an apparent honeycomb structure (Fig.~\ref{fig:fig1}c)~\cite{chen2023visualizing, jiang2023kagome}.

Doping of CoSn with Fe manifests in atomic scale modulations of the topographic height of the Co$_3$Sn surface layer (Fig.~\ref{fig:fig1}e) that can only be detected at the lattice sites of the kagome lattice but not inside the honeycomb center. This suggests the Fe atoms substitute Co atoms in the Co$_3$Sn layer ({\em c.f.}~Fig.~\ref{fig:fig1}a). The density of these atomic scale modulations continuously increases with increasing doping levels (Fig.~\ref{fig:fig1},e-g); still, even at the highest doping levels ($x=0.24$) investigated in our study, the STM topographies reveal the apparent honeycomb structure of the Co$_3$Sn surface. We do not find signatures of clustering and sub-phases in the studied islands in STM measurements, and we can detect large atomically flat terraces in topographic STM measurements on Co$_{1-x}$Fe$_x$Sn samples (Sec.~2 of Suppl. Materials). This suggests the successful formation of Co$_{1-x}$Fe$_x$Sn thin films for $0\leq x\leq 0.24$~\cite{sales2021tuning}.

The large local density of states (LDOS) of the kagome FBs of CoSn ($x=0$) manifests in a sharp peak centered at energy $E=eV\approx-200\,$meV ($V$ denotes the applied bias voltage and $e$ the electron charge) in the differential conductance ($dI/dV$) spectrum (as seen in Fig.~\ref{fig:fig1}h)~\cite{liu2020orbital, chen2023visualizing}. A comparison with results from tight-binding calculations of the surface spectral function shows that states near the $dI/dV$ peak center predominantly originate from electronic states of the two FBs near the $\Gamma$-point in the Brioullin zone, whereas states at higher energy can be attributed to electronic states of the upper flat band away from $\Gamma$ (Sec.~3.D of Suppl. Materials). In real space, the flat band LDOS is localized to the kagome honeycomb center (as seen in Fig.~\ref{fig:fig1}d) with a localization length $l\approx2-3\,$Å, whereas the spectral weight of the dispersive bands resides near the kagome lattice sites~\cite{chen2023visualizing}. 

Chemical doping of CoSn with Fe manifests as a shift of the flat band $dI/dV$ peak toward higher energy (see Fig.~\ref{fig:fig1}h). This trend is consistent with the expectation that substituting Co by Fe hole dopes the system~\cite{sales2021tuning}. A pronounced $dI/dV$ peak can be detected throughout the whole doping series. This suggests that Fe doping induces modest impurity potentials on the order of or weaker than the FB spectral width $<100\,$meV. To quantify the energetic shift $\Delta E$ of $E_{\rm F}$ induced by Fe doping, we trace the spectral position of the FB $dI/dV$ peak as a function of $x$. We detect $\Delta E>90\,$meV at $x=0.24$ (as shown in Fig.~\ref{fig:fig1}i). The energetic shift is particularly apparent at higher doping $x>0.10$, where $E_{\rm F}$ resides within the $dI/dV$ peak. Hence, the upper FB is partially filled. In the following, we examine the appearance of strongly correlated states in the partially filled FBs of Co$_{1-x}$Fe$_x$Sn as a function of Fe-doping and temperature.

\section{Nematic order in a partially-filled kagome flat band}

\begin{figure}[H]
    \centering
    \includegraphics[width=1\linewidth]{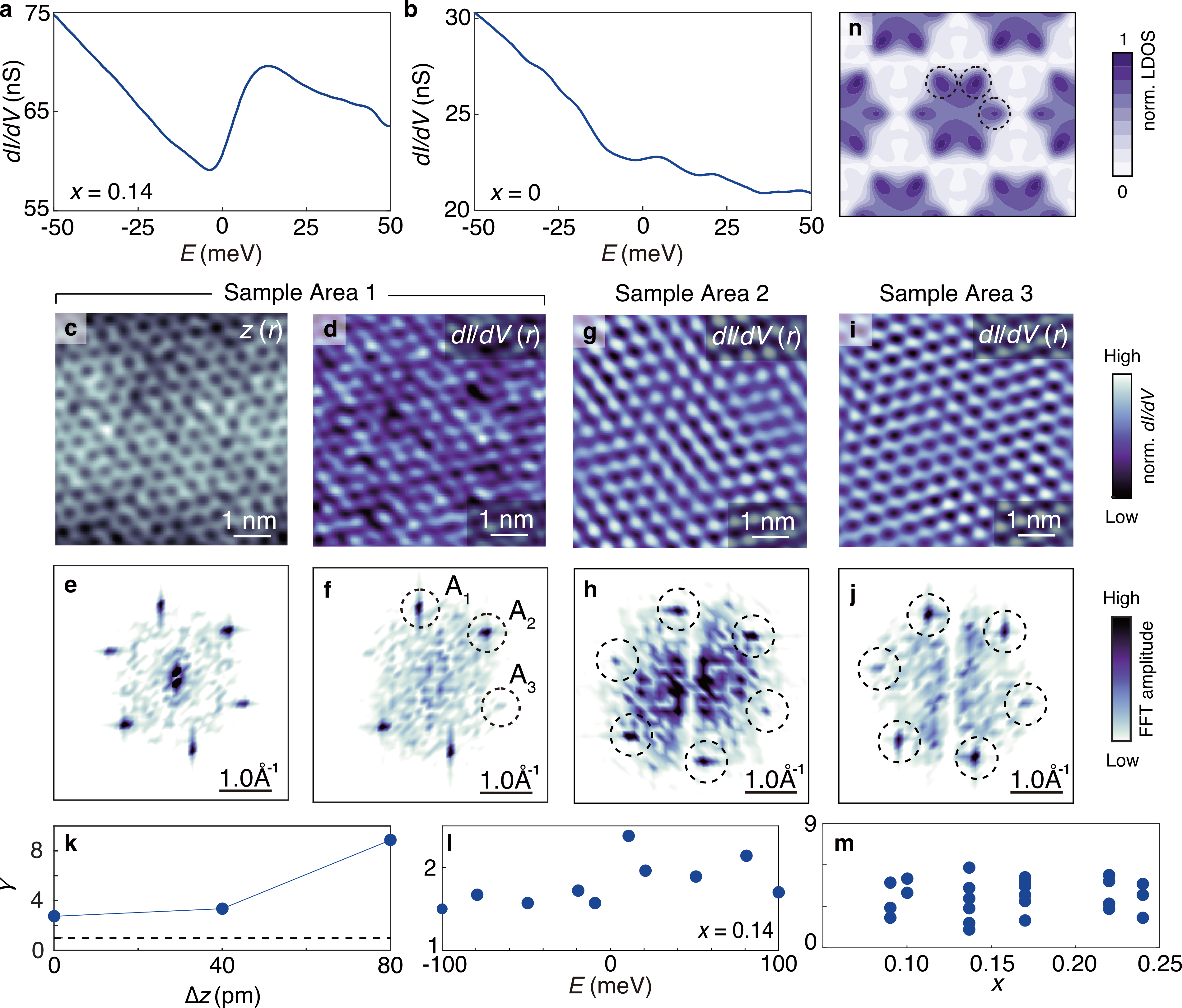}
    \caption{{\bf Nematic order parameter in the partially filled kagome flat bands.} {\bf a} and {\bf b,} $dI/dV$ spectra recorded on the surface of Co$_{0.86}$Fe$_{0.14}$Sn ($x=0.14$, $V=-50\,$mV, $I=3.0\,$nA, $V_{\rm m}=1.0\,$mV, $T=16\,$K) and CoSn, respectively ($x=0$, $V=-100\,$mV, $I=3.0\,$nA, $V_{\rm m}=1.0\,$mV). {\bf c,} STM topography of Co$_{0.86}$Fe$_{0.14}$Sn ($V=100\,$mV, $I=4.0\,$nA, $T=16\,$K). {\bf d,} $dI/dV$ map recorded in the same field of view as panel c ($V=5\,$mV, $I=4.0\,$nA, $V_{\rm m}=1.0\,$mV, $T=16\,$K, $\Delta z=40\,$pm). {\bf e} and {\bf f,} 2D-FFT maps of the data shown in panels c and d. $A_1$, $A_2$, and $A_3$ label the three sets of Bragg peaks. {\bf g} and {\bf h,} Inverse 2D-FFT maps of $dI/dV$ maps recorded at $E=0\,$meV and $E=-10\,$meV, respectively. The inverse 2D-FFT is calculated from the Bragg peaks (as highlighted by dashed circles) of the corresponding 2D-FFT shown in panels i and j. Note these maps are inverse 2D-FFT of the six Bragg peaks shown in panels i and j. The corresponding topographies and unfiltered $dI/dV$ maps are shown in Sec.~4 of the Suppl. Materials. {\bf k}-{\bf m} Shown is the dependence of the symmetry-breaking strength $\gamma$ on the reduction $\Delta z$ of the tip-sample separation, the energy $E$, and the Fe-doping level $x$. $y$-axis error bars are included within the symbols. {\bf n,} Calculated local density of states at the surface of a 50 unit cell slab of CoSn with an inter-band order parameter $\delta=180\,$meV (see Sec.~6 of the Suppl. Materials).}
    \label{fig:fig2}
\end{figure}

We first focus on moderately doped Co$_{0.86}$Fe$_{0.14}$Sn and examine the low energy electronic states at $T=16\,$K. The $dI/dV$ spectrum recorded at the honeycomb center exhibits an asymmetric dip-hump feature at $E_{\rm F}$ (Fig.~\ref{fig:fig2}a). By contrast, the $dI/dV$ spectrum recorded on undoped CoSn ($x=0$) remains featureless at $E_{\rm F}$, even at lowest temperatures $T=4\,$K (Fig.~\ref{fig:fig2}b). We carried out spectroscopic mapping experiments to examine the real space distribution of the $dI/dV$ amplitude associated with this feature. Fig.~\ref{fig:fig2}, c and d show the STM topography and $dI/dV$ map recorded at $E=5\,$meV recorded in the same field of view. Interestingly, the $dI/dV$ map breaks the six-fold rotational symmetry ($C_{6z}$) of the kagome lattice. This symmetry breaking is evident in the corresponding 2D fast Fourier transform (2D-FFT) of the $dI/dV$ map (as shown in Fig.~\ref{fig:fig2}f), in which two out of six Bragg peaks exhibit a suppressed amplitude. A closer inspection shows that the six Bragg peaks can be categorized into three subsets ($A_1$, $A_2$, and $A_3$) of pairs that have different relative amplitudes ($A_1=0.9$, $A_2=1.0$, and $A_3=0.1$). This suggests that the electronic states near $E_{\rm F}$ of the sample with $x=0.14$ Fe-doping have $C_{2z}$-symmetry. By contrast, the $dI/dV$ map recorded near $E_{\rm F}$ on undoped CoSn ($x=0$) has $C_{6z}$-symmetry (as seen in Fig.~\ref{fig:fig1}d).

We observed the $C_{6z}$-symmetry breaking characteristics in all measurements conducted on Co$_{0.86}$Fe$_{0.14}$Sn (see Sec.~4A of Suppl. Materials for additional data sets). We also find sample areas in which $dI/dV$ maps recorded near $E_{\rm F}$ exhibit a domain structure. Two examples (labelled sample areas 2 and 3), shown in Fig.~\ref{fig:fig2}, g and h, exhibit multiple nematic domains ordering along three different spatial directions. The presence of domains in the symmetry-broken low-energy electronic states rules out strain and tip effects, such as an anisotropic tip shape~\cite{da2013detection}, as the cause of the observed $C_{6z}$-symmetry breaking. The corresponding topographies (as shown in Sec.~4B of Suppl. Materials) do not exhibit crystallographic domains that could account for this effect, nor can the observed symmetry breaking be caused by a charge density wave, because the low-energy electronic states retain translation symmetry. Instead, our observations suggest that electronic correlations cause the detected $C_{6z}$ to $C_{2z}$-breaking, resulting in a nematic electronic ground state.

To obtain more experimental insight into the origin of this state, we examined the rotational symmetry breaking as a function of several experimental parameters (as shown in Fig.~\ref{fig:fig2}, k-m). For simplicity, we parameterize its strength as $\gamma=(A_1 +A_2 )/(2A_3)$; $C_{6z}$ is broken for $\gamma\neq1$. First, we varied the tip-sample separation. We find that $C_{6z}$-breaking increases when decreasing the tip-sample separation by $\Delta z$ (as seen in Fig.~\ref{fig:fig2}k). This suggests that symmetry-breaking is more prominent in electronic states with finite in-plane momentum, which have a larger out-of-plane wavefunction decay constant than those states near the crystallographic $\Gamma$ point (see Sec.~4C of Suppl. Materials for a technical discussion of this effect). We also examined $dI/dV$ maps recorded at energies away from $E_{\rm F}$. Interestingly, we detect signatures of $C_{6z}$-breaking over a wide energy range from $-100\,\text{meV}\leq E\leq100\,\text{meV}$ (as can be seen in Fig.~\ref{fig:fig2}l), and we find that $\gamma$ is largest near $E_{\rm F}$ (the energy-dependent $dI/dV$ maps are displayed in Sec.~4D of the Suppl. Materials). Finally, we studied the effect of the Fe-doping level on the $C_{6z}$-breaking. While we do not detect $C_{6z}$ to $C_{2z}$-breaking in weakly doped samples ($x=0.05$), $dI/dV$ maps recorded on samples with $x\geq0.09$ break the sixfold rotation symmetry, and this effect is found to be $x$-independent (as shown in Fig.~\ref{fig:fig2}m and Sec.~4.E of Suppl. Materials).

Given the localized character ($\approx2-3\,$Å~\cite{chen2023visualizing}) of the kagome flat band orbitals and the significant Coulomb interaction of the $3d$-orbital, it is natural to expect a symmetry-breaking order parameter that derives from strong, local density-density interactions between the FB orbitals localized at the honeycomb center within the same unit cell~\cite{chen2023visualizing}. If only one of the two FBs is involved, the mean-field (MF) symmetry-breaking Hamiltonian can be expressed as $\delta \hat H_{\rm MF} = \sum_{\vec r; \mu, \nu} \hat f^\dagger_{\vec r; \mu} \left( \vec d \cdot \vec \sigma \right)_{\mu \nu} \hat f_{\vec r; \nu}$, where $\hat f^\dagger_{\vec r; \mu}$ creates a localized electron centered around $\vec r$ with spin $\mu = \uparrow, \downarrow$, $\vec d$ is a three-dimensional real vector that parameterizes the symmetry-breaking order parameter, and $\vec \sigma$ denote the three Pauli matrices. 
This order parameter transforms as a spin-one and always breaks time-reversal symmetry. If $\vec d$ points along the $z$-direction, $C_{6z}$ is preserved, if $\vec d$ lies in the plane, no rotation symmetry remains. Neither of these scenarios is compatible with our experimental observations. 

We are thus naturally led to considering inter-band interactions between states belonging to the two FBs. First, note that the in-plane orbitals $d_{xy/ x^2 - y^2}$ are even under a mirror $M_{xy}$ which flips $z \leftrightarrow -z$, whereas the out-of-plane orbitals $d_{xz/yz}$ are odd under the same mirror. Therefore, these two sets of kagome FBs can only hybridize when spin-orbit coupling (SOC) is present. The SOC strength controls the degree of hybridization between these two orbitals, which is rather weak in the non-interacting limit. Since the energetic separation between the FB states is only on the order of a hundred milli-electron volts or less~\cite{kang2020topological}, a large enough Coulomb interaction $U$ could induce an inter-band order parameter $\delta\propto U$ that mixes the electronic states of the two FBs.

Indeed, our symmetry analysis identifies exactly one MF inter-band order parameter $\delta$ which reduces $C_{6z}$ to $C_{2z}$ (see Sec.~6 of Suppl. Materials). This order parameter hybridizes the states of the upper and lower FBs over a wide energy range $\propto\delta$ and results in a ground state with only $C_{\rm 2z}$ symmetry; the corresponding $C_{\rm 2z}$-symmetric calculated local density of states is shown in Fig.~\ref{fig:fig2}n. Band mixing is especially prominent for states near the $M$-point; in other words, the nematic ground state is predominantly composed of states with finite in-plane momentum. Additionally, the nematic order parameter increases the energy of states with finite momentum such that the mixed electronic flat band states exist over a wide energy range (as shown in Sec.~6 of Suppl. Materials). Therefore, our theoretical analysis can account for all of our experimental observations described above. This suggests the presence of inter-band nematic order, which breaks $C_{\rm 6z}$ but retains both translation and time-reversal symmetry (TRS), in the partially-filled mixed upper FB of Co$_{1-x}$Fe$_{x}$Sn over a wide Fe-doping range at $T\geq16\,$K.

\section{Temperature-driven transition into a pseudogap phase}

\begin{figure}[H]
    \centering
    \includegraphics[width=1\linewidth]{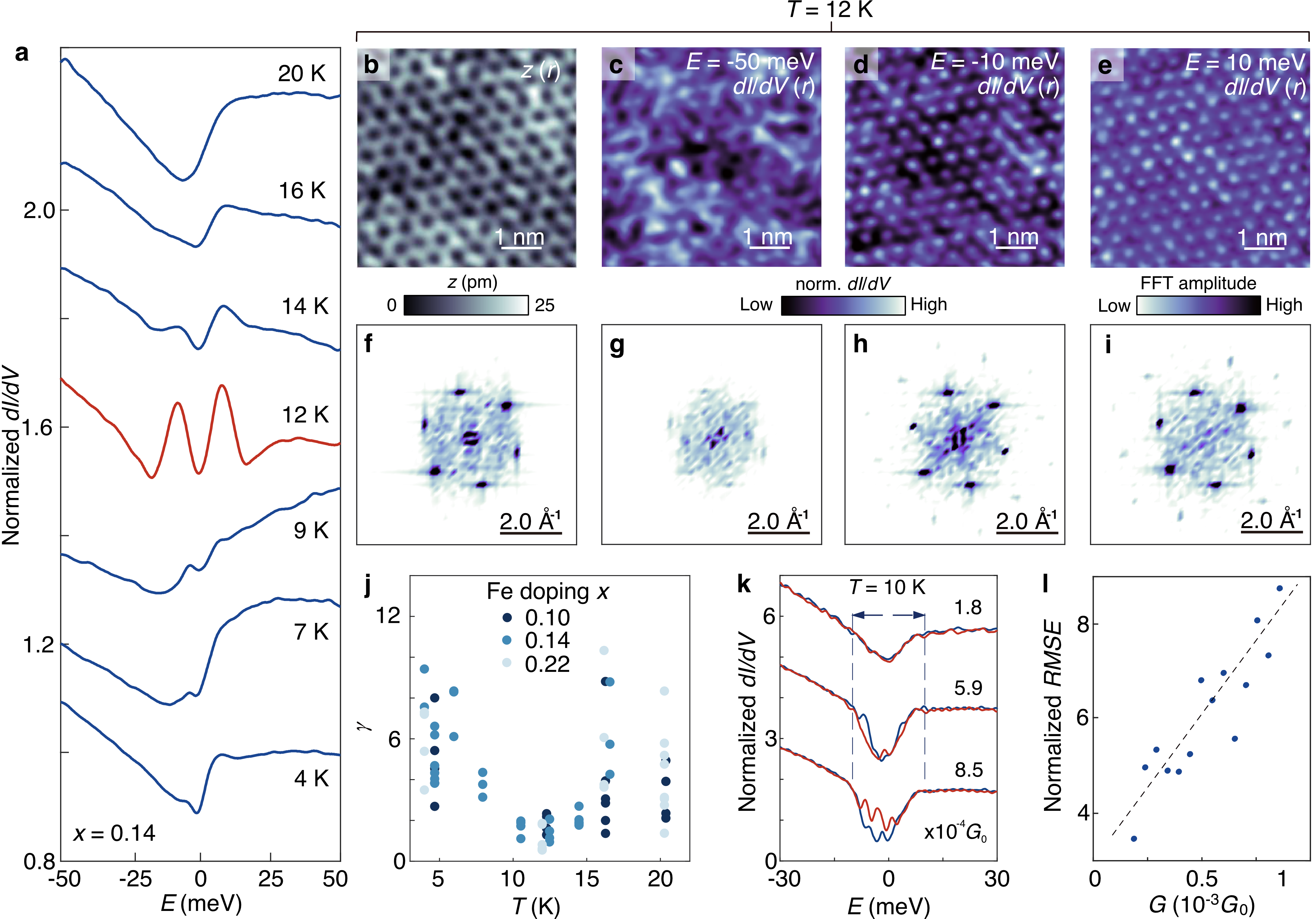}
    \caption{{\bf Observation of a pseudogap state in the partially filled kagome flat band of Co$_{0.86}$Fe$_{0.14}$Sn.} {\bf a,} Normalized $dI/dV$ spectra recorded at different indicated temperatures $T$ ($x=0.14$, $V=50\,$mV, $I=3.0\,$nA, $V_{\rm m}=1.0\,$mV). The data are vertically offset by 0.2 for clarity. {\bf b,} Representative STM topography of Co$_{0.86}$Fe$_{0.14}$Sn ($x=0.14$, $V=100\,$mV, $I=4.0\,$nA, $T=12\,$K). {\bf c}-{\bf e,} $dI/dV$ maps recorded in the same field of view as panel b at different indicated energies $E$ ($I=4.0\,$nA, $V_{\rm m}=1.0\,$mV, $T=12\,$K). {\bf f}-{\bf i,} 2D-FFT maps of the data shown in panels b-e. {\bf j,} $\gamma(T)$ near $E_{\rm F}$ plotted for samples with different indicated doping level $x$ (see Secs., 4, 8, and 11 of Suppl. Materials). Error bars are included within the symbols. {\bf k,} $dI/dV$ spectra normalized to the spectral average recorded at indicated values of the tunnel junction conductance $G$ normalized to the quantum of conductance $G_0$. Shown are the $dI/dV$ spectra of the forward (blue line) and backward (red line) voltage sweeps. The spectra are vertically offset for clarity ($V=50\,$mV, $V_{\rm m}=1.0\,$mV, $T=10\,$K). {\bf l,} Normalized root mean square error (RMSE) of the $dI/dV$ spectra shown in panel k.}
    \label{fig:fig3}
\end{figure}

Next, we studied the response of the nematic order to a lowering of the experimental temperature from 16 to 4\,K. To this end, we record $dI/dV$ spectra at the honeycomb center of the kagome lattice at different temperature set points on Co$_{0.86}$Fe$_{0.14}$Sn (Fig.~\ref{fig:fig3}a). As the temperature is lowered, the $dI/dV$ spectrum undergoes two transitions; at $T\approx14\,$K, we observe a redistribution of the spectral weight into a double peak structure. At $T\approx8\,$K, we observe another redistribution of the spectral weight that consolidates in a low-energy $dI/dV$ spectrum with marked electron-hole asymmetry at 4\,K, our experimental base temperature. These observations suggest the presence of a distinct electronic ground states in Co$_{0.86}$Fe$_{0.14}$Sn at different temperatures.

We first examine the intermediate temperature phase appearing between 8 and 16\,K. The $dI/dV$ spectrum is dominated by a set of two spectral peaks near $E_{\rm F}$. Their maxima are separated by $\Delta\approx20\,$meV and enclose a pseudogap that appears on top of a finite $dI/dV$ background. We also recorded $dI/dV$ maps at the peak positions $E=\pm 10\,$meV of the pseudogap (Fig.~\ref{fig:fig3}, d and e). The spectral weight of these $dI/dV$ peaks form a periodic real space pattern whose maxima are localized at the honeycomb center of the kagome lattice. Hence the pseudogap state arises within the kagome FB orbitals~\cite{chen2023visualizing}. By contrast, $dI/dV$ maps recorded away from $E_{\rm F}$ do not exhibit this distinct pattern (see Fig.~\ref{fig:fig3}c for $E=-50\,$meV). Additional data sets that reproduce these observations are shown in Sec.~7A of the Suppl. Materials.

The removal of spectral weight from the Fermi level (as seen in Fig.~\ref{fig:fig3}a) into the pseudogap peaks could be interpreted as a signature of electronic symmetry breaking. However, the real space distribution of the spectral weight of the $dI/dV$ peaks as seen in Fig.~\ref{fig:fig3}, d and e does not break rotation symmetry, $\gamma\approx1$ (as summarized in Fig.~\ref{fig:fig3}; also see Sec.~8 of the Suppl. Materials). Neither does it break translation symmetry, which distinguishes our observation from a charge density wave that breaks translation symmetry (e.g., as shown for the pseudogap phase of transition metal oxides in Ref.~\cite{cai2016visualizing}).

To gain further insights into the nature of this state, we investigated $dI/dV$ spectra near the transition point to the low-temperature phase at $T\approx10\,$K where the pseudogap structure is weakened (Fig.~\ref{fig:fig3}k). Interestingly, we detect significant fluctuations of the $dI/dV$ amplitude at small energies $|E|\leq10\,$meV inside the V-shaped spectral gap that increase with increasing tunnel junction conductance $G$ (as seen in Fig.~\ref{fig:fig3}l; see Sec.~7C of Suppl. Materials for the pertinent analysis.). However, these fluctuations are absent at $|E|>10\,$meV. Accordingly, $dI/dV$ maps recorded at $|E|\leq10\,$meV exhibit random noise fluctuation, while those recorded at $|E|>10\,$meV do not. Crucially, these fluctuations are absent in simultaneously recorded STM topographies (as shown in Sec.~7.D of the Suppl. Materials). Hence, we can exclude an unstable tunnel junction caused by surface adsorbates, such as mobile hydrogen atom, as the origin of these fluctuations.

While further measurements are needed to characterize the origin of the pseudogap phase, our combined experimental observations presented in Fig.~\ref{fig:fig3} suggest that it is probably related to an order parameter coupled to the charge degree of freedom that arises within the partially-filled kagome FB. In proximity to the temperature-driven phase transition into and out of the pseudogap phase, such order parameter is anticipated to exhibit significant charge fluctuations, and the process of injecting charges into this state through electron tunneling is highly sensitive to these fluctuations. Consequently, these order parameter fluctuations are expected to manifest in fluctuations of the differential conductivity as observed in our measurements (Fig.~\ref{fig:fig3}l). This understanding would also be consistent with the observed insensitivity of the pseudogap to the application of an out-of-plane magnetic field \(|B| \leq 1\,\text{T}\) (see Sec.~7.E of the Suppl. Materials).

Our temperature-dependent measurements shown in Fig.~\ref{fig:fig3}a indicate that the pseudogap phase then transitions into a low-temperature phase stabilizing at $T<8\,$K (see Sec.~9 of the Suppl. Material for a detailed characterization of this low-temperature phase). At $T=4\,$K, we detect an electron-hole asymmetric $dI/dV$ spectrum with a tall (weak) shoulder at $eV\approx10\,$meV ($eV\approx-10\,$meV) as shown in Fig.~\ref{fig:fig3}a. The corresponding $dI/dV$ maps recorded at $T=4\,$K exhibit a periodic $dI/dV$ pattern that breaks $C_{6z}$-symmetry with $\gamma\gg1$ (as summarized in Fig.~\ref{fig:fig3}j). We will later show that this low-temperature nematic phase can be qualitatively distinguished from the parent phase detected at $T>16\,$K.

\section{Response of the pseudogap phase to electron and hole doping}

\begin{figure}[H]
    \centering
    \includegraphics[width=1\linewidth]{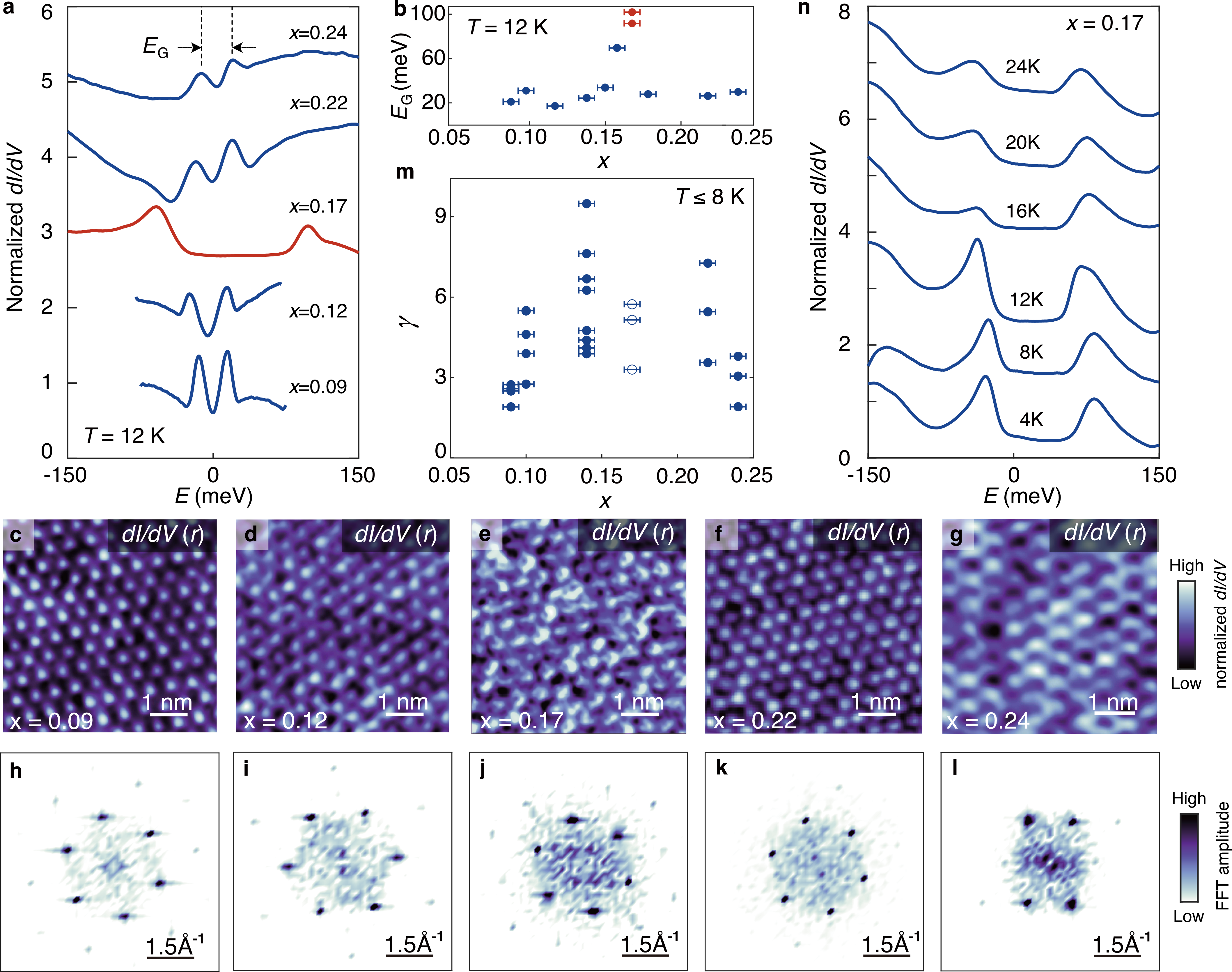}
    \caption{{\bf Response of the low-energy electronic states to Fe-doping.} {\bf a,} Normalized $dI/dV$ spectra recorded on Co$_{1-x}$Fe$_{x}$Sn samples with different indicated doping levels $x$ ($V=150\,$mV, $I=4.0\,$nA, $V_{\rm m}=1.0\,$mV, $T=12\,$K). The spectra are vertically offset by unity for clarity. {\bf b,} Spectral separation $E_{\rm G}$ of the double peaks appearing in the $dI/dV$ spectra of panel a, plotted as a function of $x$. $y$-axis error bars are included within the symbols. {\bf c}-{\bf g,} $dI/dV$ maps recorded on samples with indicated doping levels $x$ at (from left to right) $V$=10, 10, 60, 12, 15\,mV ($I=4.0\,$nA, $V_{\rm m}=1.0\,$mV, $T=12\,$K). {\bf h}-{\bf l,} 2D-FFT maps of the data shown in panels c-g. {\bf m,} Displayed is $\gamma$ as a function of $x$ extracted from 2D-FFT maps recorded near $E_{\rm F}$ at $T\leq8\,$K (See Sec.11 of Supplementary Materials). $y$-axis error bars are included within the symbols. {\bf n,} Series of normalized $dI/dV$ spectra recorded on Co$_{0.83}$Fe$_{0.17}$Sn at different indicated temperatures ($V=150\,$mV, $I=3.0\,$nA, $V_{\rm m}=1.0\,$mV). The spectra for $T\geq8\,$K are vertically offset for clarity.}
    \label{fig:fig4}
\end{figure}

The electronic ground state of a FB is known to be sensitive to the filling fraction~\cite{lee2006doping, cao2018unconventional}. To study this relation for the FBs of Fe-doped CoSn, we examined the response of the pseudogap phase to changes in the Fe-doping level. Fig.~\ref{fig:fig4}a displays representative $dI/dV$ spectra recorded at the kagome honeycomb center on samples with different Fe-doping level at 12\,K; $dI/dV$ spectra recorded on samples with other intermediate doping levels are shown in Sec.~10A of the Suppl. Materials. We find that the pseudogap at $E_{\rm F}$ can be detected between $0.09\leq x\leq 0.24$. Only at intermediate doping ($x=0.17$), we observe a significant modification of the $dI/dV$ spectrum; it exhibits two pronounced peaks separated by a large flat gap $E_{\rm G}\approx110\,$meV. The two peaks detected at $E\approx-40\,$meV and $E\approx70\,$meV, respectively are separated by a spectral gap $\approx110\,$meV, which is asymmetric about $E_{\rm F}$. Moreover, this wide-gap feature appears on top of a finite $dI/dV$ background.

We mapped out the spectral weight distribution of the peaks in the $dI/dV$ spectra for samples with different doping level by recording $dI/dV$ maps at the peak energies above $E_{\rm F}$ (as shown in Fig.~\ref{fig:fig4}, c-g). $dI/dV$ maps recorded at the peak below $E_{\rm F}$ exhibit similar characteristics and are shown in Sec.~10B of the Suppl. Materials. The localized character of the spectral weight at the honeycomb center of the kagome lattice is detected in all samples with $x\geq0.09$. This indicates that the pseudogap phase exists across a wide doping range. Moreover, the $dI/dV$ map of the $x=0.17$ sample reveals an open circular shape ('{\em bitten donut}') of the localized $dI/dV$ amplitude (Fig.~\ref{fig:fig4}d, also see Sec.~11 of the Suppl. Materials). This contrasts with the rather dot-like shape of the real space $dI/dV$ pattern found at other doping levels.

We also characterized the low temperature phase at $T<8\,$K as a function of Fe-doping level (see Sec.~12 of the Suppl. Materials for the $dI/dV$ spectra and maps). Except for the sample with $x=0.17$, the $dI/dV$ spectra recorded on samples with $x\geq0.09$ at $T<8\,$K are comparable to that of Co$_{0.86}$Fe$_{0.14}$Sn presented in Fig.~\ref{fig:fig2}a. Correspondingly, we detect rotation symmetry breaking $\gamma>1$ in $dI/dV$ maps at all Fe-doping levels $x\geq0.09$. This suggests that the nematic low temperature phase is present over a wide doping range. Most importantly, the summary of the $C_{\rm 6z}$-breaking analysis presented in Fig.~\ref{fig:fig4}m shows that $\gamma$ exhibits a non-monotonic dependence on $x$ at $T<8\,$K; it is weakest at high and low doping levels and increases toward intermediate doping levels near $x=0.17$. This observation qualitatively distinguishes the low temperature nematic phase from the the phase detected at $T\geq16\,$K for which $\gamma$ is independent of $x$ (as seen in Fig.~\ref{fig:fig2}l). It is interesting to note recent results from magnetometry measurements of Co$_{1-x}$Fe$_{x}$Sn that report evidence for antiferromagnetic interactions as well as a magnetic transition at $T<10\,$K across the same doping range $(0<x<0.2)$~\cite{sales2021tuning}. Our results suggest the high-temperature nematic phase at $T\geq16\,$K breaks $C_{\rm 6z}$ but retains TRS, whereas the low-temperature nematic phase breaks both $C_{\rm 6z}$ and TRS. As the TRS breaking is not directly observable in the experiments presented in this study, the differences in the symmetry breaking characteristics of the high and low temperature nematic phases remain an interesting open problem.

Next, we studied the response of the wide-gap feature detected in the $x=0.17$ sample to a temperature change. Fig.~\ref{fig:fig4}n displays $dI/dV$ spectra recorded at the kagome honeycomb center at different indicated temperatures. For temperatures between 4 and 12\,K, the $dI/dV$ spectrum remains qualitatively unchanged. When the temperature is increased to above 12\,K, the $dI/dV$ amplitude of the two peaks decreases and the gap softens. However, even at 24\,K, the highest temperature that we can experimentally reach, the two peaks are still visible. Hence, the temperature dependence of the low-energy electronic states of Co$_{0.83}$Fe$_{0.17}$Sn is qualitatively different from those at other doping levels (as exemplified for Co$_{0.86}$Fe$_{0.14}$Sn in~Fig.~\ref{fig:fig3}a). These observations suggest the presence of a distinct electronic ground state in Co$_{0.83}$Fe$_{0.17}$Sn across a wide temperature range.

\section{Experimental evidence for an orbital-selective Mott state}

\begin{figure}[H]
    \centering
    \includegraphics[width=1\linewidth]{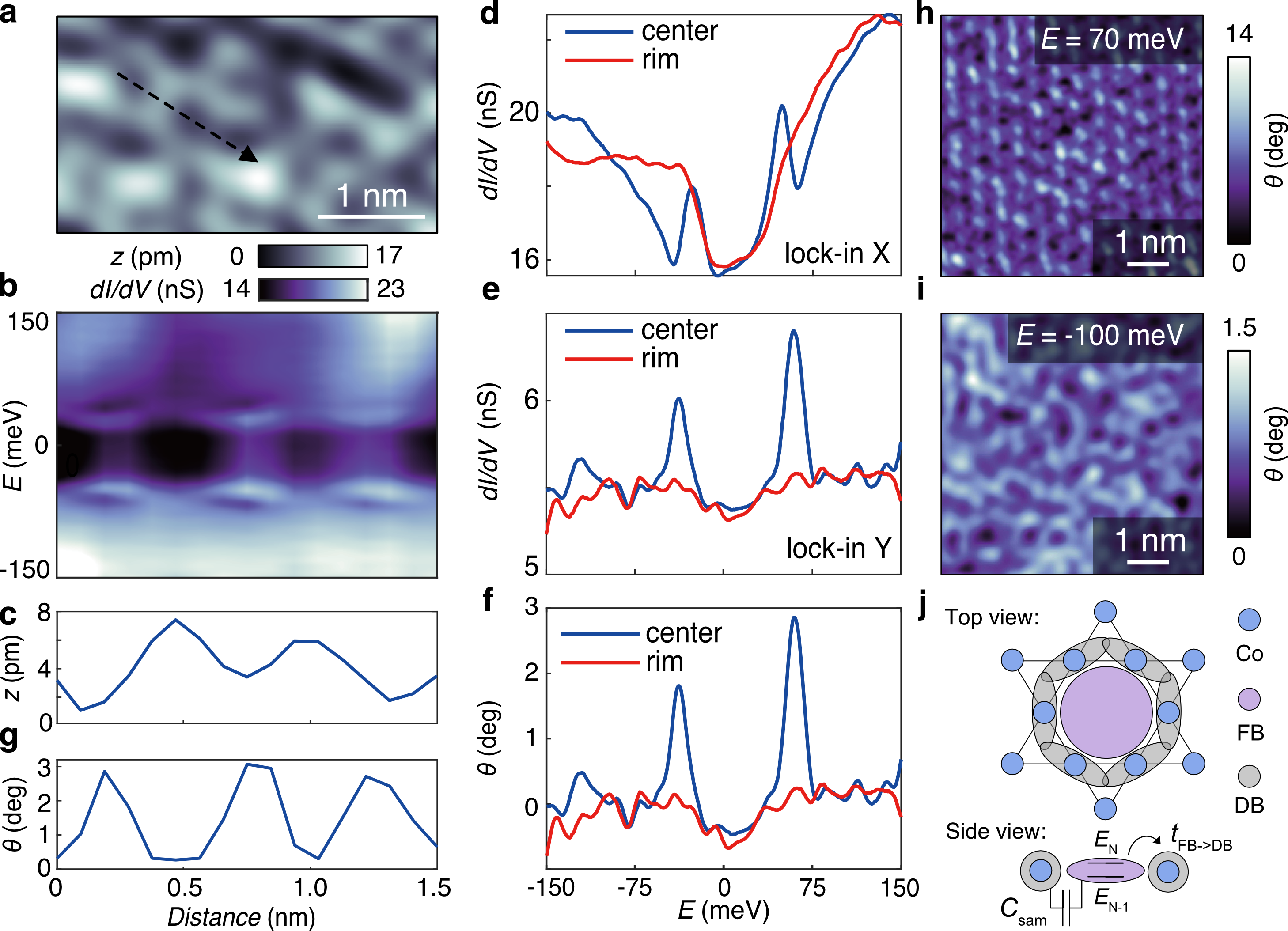}
    \caption{{\bf Experimental evidence for an orbital-selective Mott state in Co$_{0.83}$Fe$_{0.17}$Sn} {\bf a,} STM topography recorded on the surface of Co$_{0.83}$Fe$_{0.17}$Sn ($x=0.17$, $V=100\,$mV, $I=2.0\,$nA, $T=12\,$K). {\bf b} and {\bf c,} shown are a sequence of $dI/dV$ spectra and the corresponding topographic profile recorded along the black dashed arrow in panel a ($V=150\,$mV, $I=3.0\,$nA, $V_{\rm m}=1.0\,$mV, $T=12\,$K). {\bf d} and {\bf e}, $dI/dV$ spectra (X- and Y-channel of lock-in amplifier) recorded at the honeycomb center and rim positions of the kagome lattice extracted from panel b. {\bf f,} Lock-in phase $\theta$ calculated from the data shown in panels d and e. {\bf g,} $\theta$ plotted as function of the tip position along the black dashed line in panel a. {\bf h} and {\bf i,} Shown are the spatially resolved lock-in phases calculated from the in-phase and quadrature components of the measured $dI/dV$ amplitude at $E=70\,$meV and $E=-100\,$meV, respectively. {\bf j} Schematic model of the $3d$-orbital derived FB (pink color) and dispersive band (DB, grey color) orbitals on the kagome lattice of Co atoms (blue color). The hopping $t_{\rm FB->DB}$ and sample self-capacitance $C_{\rm sam}$ are indicated.}
    \label{fig:fig5}
\end{figure}

The wide spectral gap detected in Co$_{0.83}$Fe$_{0.17}$Sn (as seen in Fig.~\ref{fig:fig4}a) suggests the presence of an orbital-selective insulating state appearing in the mixed upper kagome flat band that is partially occupied. To support this hypothesis, we conducted phase-sensitive lock-in measurements on another CoSn island of the $x=0.17$ sample, which exhibits similar characteristics to those shown in Fig.~\ref{fig:fig4}. We recorded a set of $dI/dV$ spectra along a line that crosses several kagome unit cells (Fig.~\ref{fig:fig5}, a-c) and monitor the in-phase ($X$-channel) and quadrature ($Y$-channel) components of the lock-in measurement. Consistent with results from spectroscopic mapping (see~Fig.~\ref{fig:fig4}d), the double peak structure in the $dI/dV$ spectrum is most pronounced near the honeycomb center and weaker at the rim (Fig.~\ref{fig:fig5}d). We detect a change of the quadrature amplitude when the tip is located at the honeycomb center and tunneling occurs at the energies of the $dI/dV$ peaks of the wide-gap feature (as seen in Fig.~\ref{fig:fig5}e); when tunneling occurs at other energies or spatial positions, the quadrature remains unchanged. This quadrature change maps to a change of the lock-in phase $\theta=\text{atan}[(Y-Y_0)/X]$ as seen in Fig.~\ref{fig:fig5}f ($Y_0\approx5.5\,$nS denotes the lock-in Y channel background).

We mapped out this phase change in real space and calculated the lock-in phase map from the in-phase and quadrature components of the measured $dI/dV$ amplitude. The phase map recorded at the energy ($E=70\,$meV) of the $dI/dV$ peak of the wide-gap feature exhibits a periodic variation of the lock-in phase in real space (as seen in Fig.~\ref{fig:fig5}h). We detect significant phase changes (maximum amplitude $\theta\approx14$°) which are strongest at the honeycomb center of the kagome lattice while the phase remains unchanged at the rim position. We do not detect a spatially periodic phase change at energies away from the wide-gap feature ($E=-100\,$meV) as shown in Fig.~\ref{fig:fig5}i (the corresponding in-phase and quadrature maps are shown in Sec.~13 of the Suppl. Materials). This observation can be explained by the presence of an orbital selective Mott state within the mixed upper kagome flat band, which is strongly localized at the honeycomb center of the kagome lattice, as follows.

In general, when a voltage difference $V(t)= V + V_{\rm mod}\cos(\omega t)$, comprising of the the d.c. bias voltage $V$ and the a.c. lock-in voltage $V_{\rm mod}$ with frequency $\omega$, is applied to the STM junction, the current signal will contain the usual tunneling current, as well as an additional contribution caused by the charging and discharging processes of the effective capacitor formed between the STM tip and sample. The differential admittance of the circuit is given by $Y(\omega) = G(V) + i\omega C_{\rm eff}(V)$, where $G(V)$ is the bias voltage-dependent tunnel conductance related to the LDOS and $C_{\rm eff}(V)$ is the local differential capacitance. The latter can be further divided into the self capacitance of the STM tip $C_{\rm tip}$, the mutual capacitance between the tip and sample $C_{\rm mut}$, and the self capacitance of the sample $C_{\rm sam}$. They are connected sequentially leading to the total effective capacitance as $\frac{1}{C_{\rm eff}} = \frac{1}{C_{\rm tip}} + \frac{1}{C_{\rm mut}} + \frac{1}{C_{\rm sam}}$. The first two contributions are mainly determined by the STM tip geometry. The third term can change dramatically with both $V$ and the spatial position of the STM tip. When tunneling occurs into the states of an orbital selective Mott phase in the mixed upper FB,whose orbital is localized to the honeycomb center, the energy difference between the two relevant charge states is $E_{N} - E_{N - 1}\approx U (N - 1) - \mu + e V$ ($N$ and $\mu$ denote the charge occupation and chemical potential, respectively). ${C_{\rm sam}}$ of such discrete energy states vanishes for general values of $V$ but forms a delta function for special values of $V$, which cause different charging states to be degenerate ($E_N = E_{N - 1}$) or resonating. As long as $C_{\rm sam}\leq C_{\rm tip},\,C_{\rm mut}$, $C_{\rm eff}$ will peak at those resonant values of $V$ leading to a corresponding change in the lock-in phase (as modeled in detail in Sec.~13 of Suppl. Materials).

Hence, the observation of the energy and position-dependent change of the lock-in phase is consistent with the presence of an orbital-selective Mott (OSM) state~\cite{nakatsuji2000quasi, anisimov2002orbital, vojta2010orbital} in the mixed $3d$-orbital derived upper FB of Co$_{0.83}$Fe$_{0.17}$Sn that is strongly localized in real space~\cite{chen2023visualizing}. This OSM state exists in the presence of $3d_{\rm xz/yz}$ and $3d_{\rm z^2}$-orbital derived dispersive bands at the Fermi energy~\cite{kang2020topological, jiang2023kagome}, which cause the finite $dI/dV$ background (as seen in Fig.~\ref{fig:fig4}a) and drain the charge of the OSM state with rate $t_{\rm FB\rightarrow DB}$ (as schematically shown in Fig.~\ref{fig:fig5}j). Important to this interpretation is that the $p$-orbital states of Sn are gapped out to high and low energy via the selective coupling to a subset of the Co $3d$-orbitals~\cite{jiang2023kagome, chen2023visualizing}. The resonant values of $V$ in phase-sensitive measurements provide an estimate of the position of upper and lower ``Hubbard bands'' that give rise to the $dI/dV$ peaks in the $dI/dV$ spectrum (as seen in Fig.~\ref{fig:fig4}f). The extracted $U'=E_{\rm G}\approx110\,$meV suggests a sizeable renormalization of the orbital $U\approx5-6\,$eV~\cite{kang2020topological} owing to electrostatic screening through the itinerant electrons, while the asymmetry about Fermi level indicates a finite amount of hole-doping within the OSM state~\cite{zaanen1985band}.

\section{Strongly Correlated Phase Diagram of a Partially Filled Kagome Flat Band}

Our experiments on the doping sequence of Co$_{1-x}$Fe$_x$Sn ($0\leq x\leq0.24$) reveal that nematic order and the pseudogap phase emerge only with Fe doping $x>0.05$ (see Sec.~5 of Suppl. Materials). This suggests weak electron correlations when the upper flat band is slightly hole-doped ($x \leq 0.05$). The orbital-selective insulator, detected in samples with $x = 0.17$, does not occur in a stoichiometric compound, posing challenges in determining the exact electron filling of the upper flat band in our STM experiments. Nevertheless, our observations of the pseudogap phase at $x<0.17$ and $x>0.17$ along with the dome-like behavior of the nematicity parameter $\gamma$ at $T < 8 K$ (see Fig.~\ref{fig:fig4}m) align with the expected phenomenology of a Mott state in a half-filled band~\cite{lee2006doping}.

Moreover, in a simplistic rigid-band model (as shown in Suppl. Sect.~3), the Fermi level is expected to shift by $\Delta E' \approx 40\,\text{meV}$ at $x=0.17$ compared to undoped CoSn. This shift is similar in magnitude to the experimentally detected shift of the flat band $dI/dV$ peak maximum $\approx 70\,\text{meV}$ ({\em c.f.}~Fig.~\ref{fig:fig1}i). Crucially, due to the nematic symmetry breaking, the hole doping in the upper flat band is not expected to depend linearly on the chemical doping $x$. The nematic order parameter $\delta \propto U'$ increases the energy of the upper flat band states, especially near the $K$- and $M$-points (as shown in Sec.~6 of Suppl. Materials). This mechanism could amplify the effect of chemical Fe-doping in the mixed upper flat band toward realizing a half-filled FB in Co$_{0.83}$Fe$_{0.17}$Sn.

Hence, our experimental observations across a wide Fe-doping ($0\leq x\leq 0.24$) and temperature ($4\,\text{K}<T\leq25\,\text{K}$) range establish a rich phase diagram of strongly correlated states in the partially filled kagome FBs of Co$_{1-x}$Fe$_{x}$Sn (as shown in Fig.~\ref{fig:fig1}b; also see Sec.~14 of the Suppl. Materials for an overview of the investigated temperatures and Fe-doping levels). Our experimental observations suggest the presence of a nematic order parameter, a pseudogap phase, and an OSM state caused by the complex interplay between the orbital and charge degrees of freedom at different flat band occupations and temperatures. Future experiments, using different sets of experimental probes, will be valuable to characterize each of the observed phases as well as the transitions between them in even more detail. They could also benefit from reduced disorder in modulation-doped samples by conducting atomic layer epitaxy of Fe-doped CoSn~\cite{cheng2022atomic}.

It is instructive to contextualize our observations with previously studied FB materials. In typical transition metal oxides, the odd electron filling on the metal ion combines with the strong on-site interaction due to the localized nature of $d$-orbitals to realize a strongly correlated phase diagram that is often understood in the context of a doped Mott insulator~\cite{lee2006doping}. By contrast, the FBs in Co$_{1-x}$Fe$_x$Sn originate from the geometric frustration of the kagome lattice~\cite{kang2020topological, meier2020flat, liu2020orbital} whose combination with the $3d$-orbital degeneracy and strong Coulomb interactions ($U'>100\,$meV) results in a qualitatively distinct correlated phase diagram. At high temperatures, the phase diagram is dominated by the inter-band nematic order over a wide doping range ($0.09\leq x\leq0.24$, as shown in Fig.~\ref{fig:fig2}l), and we interpret it as the parent phase of the phase diagram shown in Fig.~\ref{fig:fig1}b. For example, the additional orbital degree of freedom in the mixed upper flat band, imparted by the nematic order parameter, manifests in the symmetry-breaking characteristics ('{\em bitten donut}') of the wave function of the OSM state in Co$_{0.83}$Fe$_{0.17}$Sn (as shown in Fig.~\ref{fig:fig4}e). Such characteristics would be absent without nematic order in a single band picture. Inter-band mixing mediated by strong local density interactions between FB electrons, as discovered in this work, should be ubiquitous in transition metal-based kagome materials owing to their $d$-orbital degeneracy. Hence, this material class could offer a novel platform to study nematic order emerging from strong electronic correlations at $W<U$.

An OSM state was initially discussed in multi-band transition-metal oxides~\cite{nakatsuji2000quasi} and iron chalcogenides~\cite{yi2013observation}. Our results from spectroscopic STM measurements, in particular the orbital-selective insulating character detected in phase-sensitive lock-in measurements, strongly suggest the presence of this state in Co$_{1-x}$Fe$_x$Sn, enabled by the $3d$-orbital degeneracy of Co. While model-calculations would be desirable to further strengthen this interpretation, Fe-doped CoSn could offer opportunities to shed light on the nature of the OSM state, as well as the response of this state to both electron and hole doping into the pseudogap phase (as seen in Fig.~\ref{fig:fig4}a). Theoretical analyses predict an OSM state is separated from the metallic side of the spectrum by a quantum critical point~\cite{senthil2008critical, vojta2010orbital}. Hence, the FBs of Co$_{1-x}$Fe$_{x}$Sn could provide new avenues to examine the appearance of a critical Fermi surface~\cite{senthil2008critical} at the OSM transition as a function of Fe-doping near $x=0.17$.

On the other hand, the kagome FBs of CoSn are predicted to possess a nontrivial Z$_2$ invariant in presence of SOC~\cite{kang2020topological}, and this nontrivial topology remains intact in the presence of the nematic order parameter (as discussed Section~6 of the Supplementary Materials). Hence, the kagome FBs share the intrinsic topological character of the FBs of moiré materials~\cite{song2019all, wu2019topological}, which themselves exhibit a wide range of interacting many-body states~\cite{ mak2022semiconductor, nuckolls2024microscopic}. Our observations suggest that Co$_{1-x}$Fe$_x$Sn could be a promising platform for exploring $3d$-orbital induced strongly correlated states emerging in a Z$_2$ topological flat band, complementing recent advances on many-body states appearing in the topological Chern bands of moiré materials~\cite{regnault2011fractional, okamoto2022topological, zeng2023thermodynamic, park2023observation, xu2023observation}. More broadly, the results presented in this work establish the exciting prospects of controlling the filling of $3d$-orbital derived kagome FBs via chemical doping toward discovering novel quantum phases of matter.

\clearpage
\section{Figures}

\newpage

\clearpage

\clearpage
\newpage

\newpage
\section{Methods}

\normalsize\textbf {Molecular beam epitaxy of CoFeSn thin films}\\
The Co$_{1-x}$Fe$_x$Sn thin films presented in this study were prepared using a home-built molecular beam epitaxy system. Details of the film deposition of CoSn films and substrate preparation procedures are described elsewhere~\cite{chen2023visualizing}. Films of nominal thickness 50\,nm were deposited on Nb-doped SrTiO$_3$(111) single crystals (CrysTec, dimensions: 5$\times$5$\times$0.5\,mm$^3$, Nb concentration: 0.05\,wt\%) by co-evaporating Co, Fe, and Sn from thermal effusion cells. The beam equivalent pressures of the Co and Fe effusion cells were adjusted to target the nominal Fe doping ratio $x$. The actual doping ratio was determined using X-ray diffraction measurements (see Sec.~1 of the Supplementary Materials). The X-ray diffraction measurements were conducted after the STM measurements on the respective sample.

\normalsize\textbf {Scanning Tunneling Microscopy (STM) Measurements}\\
The as-grown Co$_{1-x}$Fe$_x$Sn thin films were {\em in-situ} transferred to a home-built STM system without breaking the ultra-high vacuum (UHV). No post-annealing of the thin films after growth was conducted. STM measurements were performed at cryogenic temperatures (4\,K$\leq T\leq24\,$K) and UHV conditions ($p\leq1.4\times10^{-10}\,$mbar) using a chemically etched tungsten STM tip. The tip was prepared on a Cu(111) surface by field emission and controlled indention and calibrated against the Cu(111) Shockley surface state before each set of measurements. $dI/dV$ spectra were recorded using standard lock-in techniques with a small bias modulation $V_{\rm m}$ chosen between $1\,\text{mV}\leq V_{\rm m}\leq10\,\text{mV}$ at a frequency $f=3.971\,$kHz. All $dI/dV$ maps were recorded using the multi-pass mode to avoid set-point effects. The set-point tunnel current $I$, bias voltage $V$, lock-in modulation $V_{\rm m}$, and temperature $T$ of each measurement are indicated in the respective figure captions.

\bibliography{bibliography}

\section{Acknowledgments}

We gratefully acknowledge valuable discussions with Peter Wahl, B. Andrei Bernevig, Kam Tuen Law, and Raquel Queiroz. This work was primarily supported by the Hong Kong RGC (Grant Nos. 26304221 and C6033-22G) and the Croucher Foundation (Grant No. CIA22SC02) awarded to B.J. Y.H., X.Y. and H.C.P. acknowledge support of the National Key R\&D Program of China (Grants No.2021YFA1401500), the Hong Kong RGC (Grant No. 26308021), and the Croucher Foundation (Grant No. CF21SC01). C.C. acknowledges support from the Tin Ka Ping Foundation.

\section{Author Contributions}

B.J., C.C., and J.Z. designed the experiment. J.Z. and C.C. grew the thin film samples and conducted the scanning tunneling microscopy measurements. C.C., J.Z., Y.W., and L.L. analyzed the experimental data. Y.H. performed the model calculations with the help from X.Y. B.J. and H.C.P. supervised the study. All authors discussed the results and contributed to the writing of the manuscript.

\section{Competing Interest Declaration} The authors declare no competing financial interest.

\section{Data Availability Statement}
Replication data for ’Cascade of strongly correlated quantum states in a partially filled kagome flat band’ can be accessed at http://dx.doi.org/10.5281/zenodo.11089937 (2024).

\end{document}